\documentclass[5p]{elsarticle}

\usepackage{lineno,hyperref}
\modulolinenumbers[5]

\journal{}

\usepackage{graphicx}   
\usepackage{latexsym}   
\usepackage{enumerate}
\usepackage{color}

\begin{document}
\begin{frontmatter}

\title{Nuclear interactions and net-proton number fluctuations\\ in heavy ion collisions at the SIS18 accelerator}
\author{Jan~Steinheimer$^1$, Yongjia~Wang$^2$, Ayon Mukherjee$^{1,6}$, Yunxiao~Ye$^{2,3}$, Chenchen~Guo$^4$, Qingfeng~Li$^{2,5}$ and Horst~Stoecker$^{1,6,7}$}
\address{$^1$ Frankfurt Institute for Advanced Studies, Ruth-Moufang-Str. 1, D-60438 Frankfurt am Main, Germany\\
$^2$ School of Science, Huzhou University, Huzhou 313000, China\\
$^3$ Department of Physics, Zhejiang University, Hangzhou 310027, China\\
$^4$ Sino-French Institute of Nuclear Engineering and Technology, Sun Yat-sen University, Zhuhai 519082, China\\
$^5$ Institute of Modern Physics, Chinese Academy of Sciences, Lanzhou 730000, China\\
$^6$ Institut f\"ur Theoretische Physik, Goethe-Universit\"at Frankfurt, Max-von-Laue-Strasse 1, D-60438 Frankfurt am Main, Germany\\
$^7$ GSI Helmholtzzentrum f\"ur Schwerionenforschung GmbH, D-64291 Darmstadt, Germany}


\begin{abstract}
The effect of nuclear interactions on measurable net-proton number fluctuations in heavy ion collisions at the SIS18/GSI accelerator is investigated. The state of the art UrQMD model including interaction potentials is employed. It is found that the nuclear forces enhance the baryon number cumulants, as predicted from grand canonical thermodynamical models. The effect however is smeared out for proton number fluctuations due to iso-spin randomization and global baryon number conservation, which decreases the cumulant ratios. For a rapidity acceptance window larger than $\Delta y> 0.4$ the effects of global baryon number conservation dominate and all cumulant ratios are significantly smaller than 1.
\end{abstract}


\end{frontmatter}


One of the most important remaining questions in the field of high energy heavy ion physics is to determine the onset of deconfinement \cite{Stephanov:1998dy,McLerran:2007qj,Alford:2007xm}. 
To this goal large scale experiments are and will be performed in several accelerator facilities worldwide, e.g. at the SPS at CERN, the RHIC, FAIR, NICA and JPARC. Here heavy ions are collided with increasing beam energy. Through the study of the beam energy dependence of various observables one hopes to find unambiguous signals for the appearance of a phase transformation of hadrons to their quark and gluon constituents. 

At the highest beam energies and lowest net-baryon densities this transformation proceeds as a smooth crossover as predicted by lattice QCD simulations \cite{Borsanyi:2010cj,Bazavov:2010sb}. At higher values of the net-baryon density lattice QCD methods are not applicable and one is left with effective models. As these models are often not sufficiently constraint, the predictions for the nature of the QCD phase transition vary significantly over model-space. Consequently it would be very important if the order of the phase transition at large density could be verified experimentally.

Recently a focus on possible observables for a phase transition and the associated critical endpoint is on fluctuations of conserved charges, e.g. the net-baryon number, strangeness and electric charge \cite{Gupta:2011wh,Luo:2011rg,Luo:2017faz,Herold:2016uvv,Zhou:2012ay,Wang:2012jr,Karsch:2011gg,Schaefer:2011ex,Chen:2011am,Fu:2009wy,Cheng:2008zh}. In a grand canonical thermodynamic ensemble the cumulants of the net-charge distribution functions should diverge at the critical point of the phase transition, due to the divergence of the correlation length.
It was therefore suggested, that the measurement of the net-proton number fluctuations in a fixed rapidity interval could reveal the onset of deconfinement and/or the critical endpoint of QCD. The measured rapidity interval has to be much smaller than the total systems rapidity width and larger than the correlation length \cite{Jeon:2000wg,Koch:2008ia}.
However, the system created in heavy ion collisions can hardly be treated as a grand canonical system in thermal equilibrium, thus the measured cumulants are also affected by other aspects of the dynamical evolution, many of which have been discussed in recent literature \cite{Bzdak:2012ab,Bzdak:2016qdc,Kitazawa:2016awu,Feckova:2015qza,Begun:2004gs,Bzdak:2012an,Gorenstein:2008et,Gorenstein:2011vq,Sangaline:2015bma,Tarnowsky:2012vu,Xu:2014jsa,Adamczyk:2014fia,Adamczyk:2013dal}.
The systems created in these nuclear collisions are very small, rapidly expanding and therefore a detailed understanding and interpretation of the measured moments is difficult due to uncertainties in the centrality determination, efficiency corrections and acceptance cuts.
To address these experimental uncertainties one employs models to simulate the dynamical expansion of the system created in the heavy ion collision. There are two main approaches, a fluid dynamical description and/or a microscopic transport description. The fluid dynamical description has the advantage that any equation of state can be easily introduced and effects of spinodal decomposition due to a phase transition can be described in a controlled manner \cite{Steinheimer:2012gc,Chomaz:2003dz,Randrup:2003mu,Sasaki:2007db}. On the other hand, thermal fluctuations, which are an important ingredient for the formation of critical fluctuations near the critical endpoint \cite{Stephanov:1998dy,Stephanov:2008qz,Koch:2008ia}, as well as the production of discrete particles from the fluid \cite{Steinheimer:2017dpb} cannot be easily introduced.

Alternatively one can use a microscopic transport model, which naturally includes thermal fluctuations and usually describes the evolution of discrete particles. On the downside it is very challenging to introduce the dynamics of a phase transition and critical point in such a transport approach. Especially a change of the effective degrees of freedom, as expected at the deconfinement transition, is not easily introduced in a consistent manner.

In this paper we will address another important contribution to measured proton number fluctuations. In previous work with grand canonical models of dense QCD it was suggested that the interactions of nucleons can have a significant impact on the measured net-proton cumulants at low ($\sqrt{s_{\mathrm{NN}}} < 20$ GeV) beam energies \cite{Fukushima:2014lfa,Vovchenko:2015pya,Vovchenko:2016rkn,Mukherjee:2016nhb}. In the following we will investigate how important the effect of the nuclear interactions on the measured proton number fluctuations really is within a microscopic transport approach. A similar study, but at a higher beam energy ($\sqrt{s_{\mathrm{NN}}}= 5$ GeV) came to the conclusion that no effect of the nuclear interactions could be found \cite{He:2016uei}. We will study this effect for fixed target experiments at a beam energy of $E_{\mathrm{lab}}= 1.23$ GeV/nucleon which corresponds to the current SIS18 HADES experiment. Here the effect of nuclear interactions should be much stronger and therefore is more likely to be observed.   

\section{The UrQMD model}
We will employ the microscopic transport model UrQMD in its latest version (v3.4). The UrQMD model is based on binary elastic and inelastic scattering of hadrons, at the beam energy under investigation mainly dominated by resonance excitations. It includes more than 50 different baryonic states with its anti-particles as well as 40 mesonic states. All these hadrons scatter according to measured cross sections, where available \cite{Patrignani:2016xqp}. If no data is available the cross sections are estimated using effective models or the additive quark model. If only elastic and inelastic scatterings are taken into account the model is used in its so-called cascade version. This version is able to basically describe particle multiplicities and collective motion over a wide range of beam energies \cite{Petersen:2006vm}. 
Since UrQMD is a miscroscopic transport model, global as well as local conservation of energy, momentum as well as all quantum charges (like the baryon number and electric charge) is observed.

\subsection{Adding Nuclear Potentials}
Nuclear interactions have been introduced to the UrQMD model already some time ago \cite{Bass:1998ca,Li:2005gfa}. In the case when nuclear interactions are taken into account, each hadron is represented by Gaussian wave packet with the width parameter $L$ in phase space~\cite{Bass:1998ca}. The Wigner distribution function $f_i$ of the hadron $i$ reads
\begin{equation}\label{eq1}
  f_{i}({\bf r},{\bf p},t)=\frac{1}{(\pi \hbar)^{3}}e^{-\frac{[{\bf r}-{\bf r}_{i}(t)]^{2}}{2L}}e^{-\frac{[{\bf p}-{\bf p}_{i}(t)]^2 \cdot 2L}{\hbar^2}},
\end{equation}
where $L=2 $~fm$^2$ is usually chosen for simulating collision with heavy nuclei like Au. $\textbf{r}_i$ and $\textbf{p}_i$ are the centroids of coordinate and momentum of hadron $i$, respectively. The equations of motion for $\textbf{r}_i$ and $\textbf{p}_i$ read
\begin{eqnarray}
\dot{\textbf{r}}_{i}=\frac{\partial  \langle H  \rangle}{\partial\textbf{p}_{i}},
\dot{\textbf{p}}_{i}=-\frac{\partial  \langle H \rangle}{\partial \textbf{r}_{i}}.
\end{eqnarray}
Here, {\it $\langle H \rangle$} is the total Hamiltonian function of the system, it comprises the kinetic energy and the effective interaction potential energy. The importance of the mean field potential for describing HICs has been extensively studied \cite{Stoecker:1986ci,Bertsch:1988ik}. For studying HICs at intermediate energies, the following density and momentum dependent potential has been widely used in QMD-like models \cite{Aichelin:1991xy,Hartnack:1997ez,Li:2005gfa},
\begin{eqnarray}\label{eq2}
U&=&\alpha (\rho/\rho_0)+\beta (\rho/\rho_0)^{\gamma} \nonumber \\
&+& t_{md} \ln^2[1+a_{md}(\textbf{p}_{i}-\textbf{p}_{j})^2]\rho/\rho_0.
\end{eqnarray}

Here $\alpha$=-393 MeV, $\beta$=320 MeV, $\gamma$=1.14, $t_{md}$=1.57 MeV, and $a_{md}=500$ GeV$^{-2}$ are chosen, which yields the incompressibility $K_0$=200 MeV for isospin symmetric nuclear matter. This set of parameters does give a good description of the azimuthal correlations of charged particles (the so called $v_n$) at SIS18 beam energy range \cite{Hillmann:2018nmd}. In recent years, in order to follow present progress on determining the nuclear symmetry energy and better describe the recent experimental data for HICs at SIS energies, the surface and surface asymmetry terms, as well as the bulk symmetry energy term obtained from the Skyrme potential energy density functional have been further considered in the present version. Details about these terms can be found in Refs.~\cite{Wang:2013wca,Wang:2014aba}, since they are expected to be less important for bulk properties of HICs, the chosen of those parameters will not significantly influence our results. Besides the nuclear potential, the Coulomb potential for all charged particles is also taken into account. It has been further found that with an appropriate choice of the in-medium elastic nucleon-nucleon cross section, some recent published experimental data, especially the collective flows of light clusters, can be reproduced reasonably well. See Refs.~\cite{Wang:2013wca,wyj-sym,Wang:2014aba} for more details.

\begin{figure}[t]	
\includegraphics[width=0.5\textwidth]{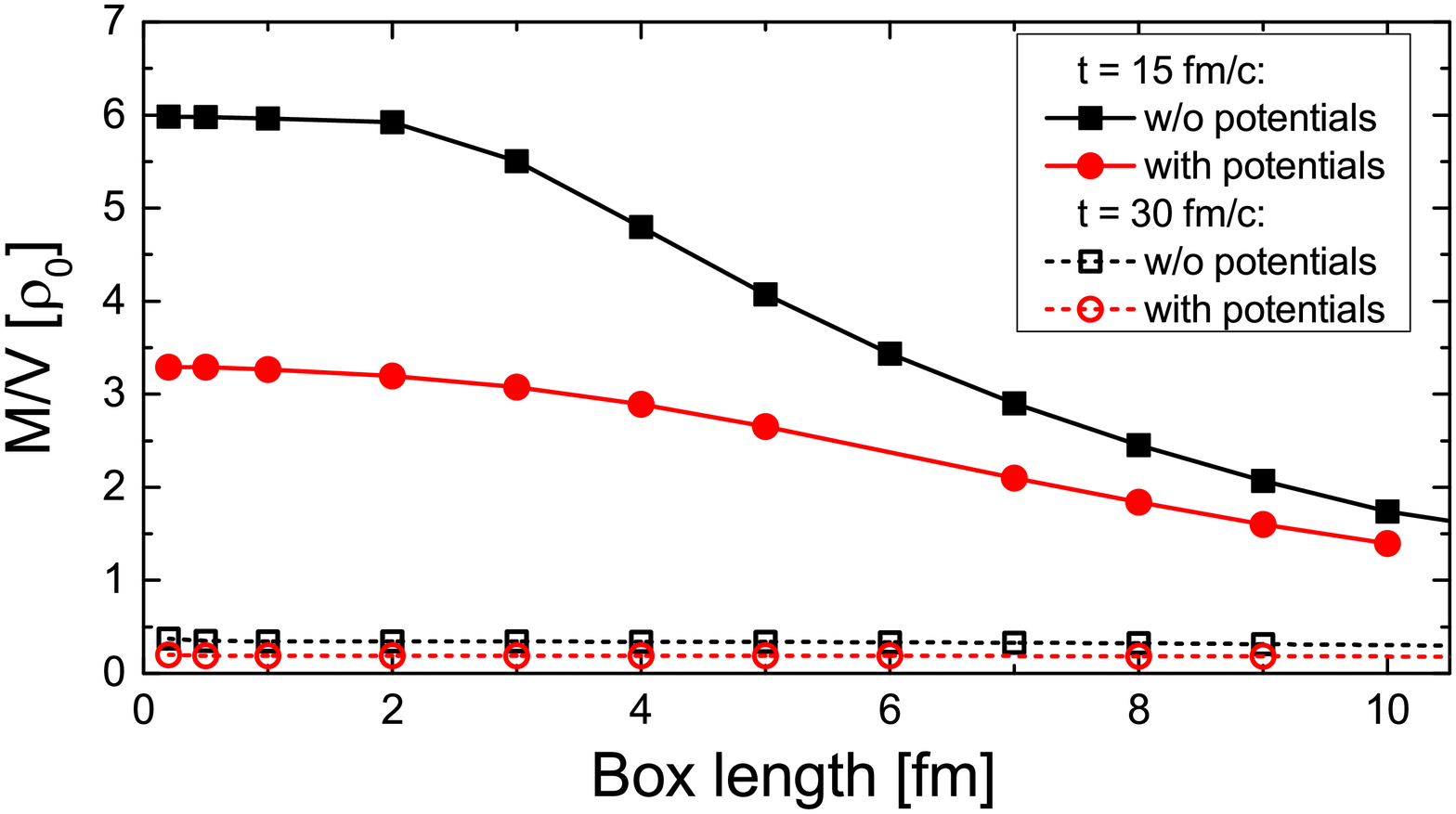}
\caption{[Color online] Baryon number density in a cubic box with length $l$. The box is centered at the center of mass of the collisions, i.e. $x=y=z=0$. Two time steps are compared, t=15 and 30 fm/c. The density is given in units of the nuclear ground state density $\rho_0 \approx 0.16 \mathrm{ fm}^{-3}$}\label{f1}
\end{figure}		

\section{Method}\label{method}

In the following, results for head-on (b=0 fm) Au+Au collisions at a fixed target beam energy of $E_{\mathrm{lab}}=1.23$A~GeV, with the UrQMD model, will be presented. In particular we will compare results where the model is used in its cascade mode with results where the long range nuclear interactions are explicitly taken into account. Note that we will treat all baryons as free baryons, i.e. they are not bound in nuclear clusters. In general this is not true and one usually applies an afterburner to calculate the cluster abundances on an event-by-event basis \cite{Li:2016wkb,Li:2016mqd}. Using such an afterburner it would be very interesting to study the effect of nuclear clustering on the baryon number fluctuations in more detail \cite{Feckova:2015qza}. However in this paper we will focus only on the effect of potential interactions and leave the cluster study to future publications. 
The importance of taking into account the nuclear interactions can already be observed from the time evolution of the baryon number density. Figure \ref{f1} shows the average net-baryon number density in a box, centered around the collision point $x=y=z=0$ with a given length $l$, at two different times $t$. The times are chosen to correspond to the time of largest compression $t\approx 15$ fm/c and the time at which inelastic processes cease $t\approx 30$ fm/c. Note that we have chosen to treat baryons as point like particles to calculate the average density in the box as there should only be integer numbers of baryon in a given volume for a single event. We have also neglected the effect that baryons may coalesce and form nuclei at a late time which will influence the extracted cumulants \cite{Feckova:2015qza}. This effect will be studied in a forthcoming paper in more detail.

It can be clearly observed that the compression in the case of the cascade version is larger than in the case where nuclear potentials are taken into account. This is mainly due to the repulsive nature of the nuclear interaction at high density.

\begin{figure}[t]	
\includegraphics[width=0.5\textwidth]{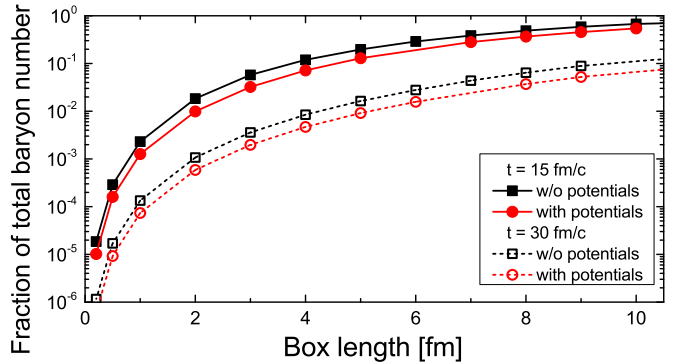}
\caption{[Color online] Fraction of baryons in the cubic box of length $l$ (see text for description). At an early time t=15 fm/c, for a box length of 10 fm about 70-80$\%$ of the baryons are enclosed within the box. At later times only about 10$\%$ are in the same box, due to the rapid expansion of the system.   
}\label{f2}
\end{figure}

In addition, figure \ref{f2} shows the total fraction of baryons within the described box. Since the total number of baryons is conserved to be 394 this fraction must be between 0 and 1. At early times it varies between 0$\%$ and 80$\%$, while at late times at most 10$\%$ of all the baryons are in the box of length 10 fm. This ratio will become important later on, in the discussion of baryon number fluctuations as effects of global baryon conservation laws are important. These effects should depend on the fraction of the total baryon number in a given acceptance/box.

In order to quantify a possible enhancement of fluctuations one usually studies ratios of cumulants of the multiplicity distributions. This is done, because the cumulants $C_n$ depend explicitely on the volume and therefore effects of the total volume cancel when the ratio is taken. Furthermore, for a Poisson distribution all cumulant ratios will be unity and the cumulant ratios for a Binomial distribution are also well known. The cumulants in the following will be defined as:

\begin{eqnarray}
C_1 &=& M  = \left\langle \mathrm{N} \right\rangle \\
C_2 &=&\sigma^2 = \left\langle (\delta \mathrm{N})^2 \right\rangle \\
C_3 &=& S \sigma^{3} =\left\langle (\delta \mathrm{N})^3 \right\rangle \\
C_4 &=& \kappa \sigma^{4} =\left\langle (\delta \mathrm{N})^4 \right\rangle - 3 \left\langle (\delta \mathrm{N})^2 \right\rangle^2 
\end{eqnarray}
 where $\delta \mathrm{N}= \mathrm{N}-\left\langle \mathrm{N} \right\rangle$ with N being the net-proton or net-baryon number in a given acceptance for a single event and the brackets denote an event average. Here $M$ is the Mean, $\sigma^2$ the variance, $S$ the Skewness and $\kappa$ the Kurtosis of the underlying multiplicity distribution.\\
 Usually one takes the following appropriate ratios of these cumulants:
 \begin{eqnarray}
C_2/C_1 &=& \sigma^2/M \\
C_3/C_2 &=&S \sigma\\
C_4/C_2 &=& \kappa \sigma^{2} 
\end{eqnarray}

The statistical errors in our simulations are estimated according to the delta-theorem \cite{Luo:2011tp}. The errors of the cumulant ratios then are:
\begin{equation}\label{delth}
error(C_r/C_2) \propto \sigma^{r-2}/\sqrt{n}~,
\end{equation}
where $n$ is the number of events and $\sigma^2= C_2$ the variance of the observable.

For reference we also cite the corresponding cumulant ratios for a Binomial distribution, which would be the correct description of uncorrelated baryons where the total baryon number is conserved globally.

 \begin{eqnarray}\label{bi}
{C_2/C_1}^{Binomial} &=& 1-p \nonumber \\
{C_3/C_2}^{Binomial} &=& 1-2p \nonumber \\
{C_4/C_2}^{Binomial} &=& 1-6p(1-p)
\end{eqnarray}
where $p$ is the fraction of the total baryon number within a given acceptance/box.

\begin{figure}[t]	
\includegraphics[width=0.5\textwidth]{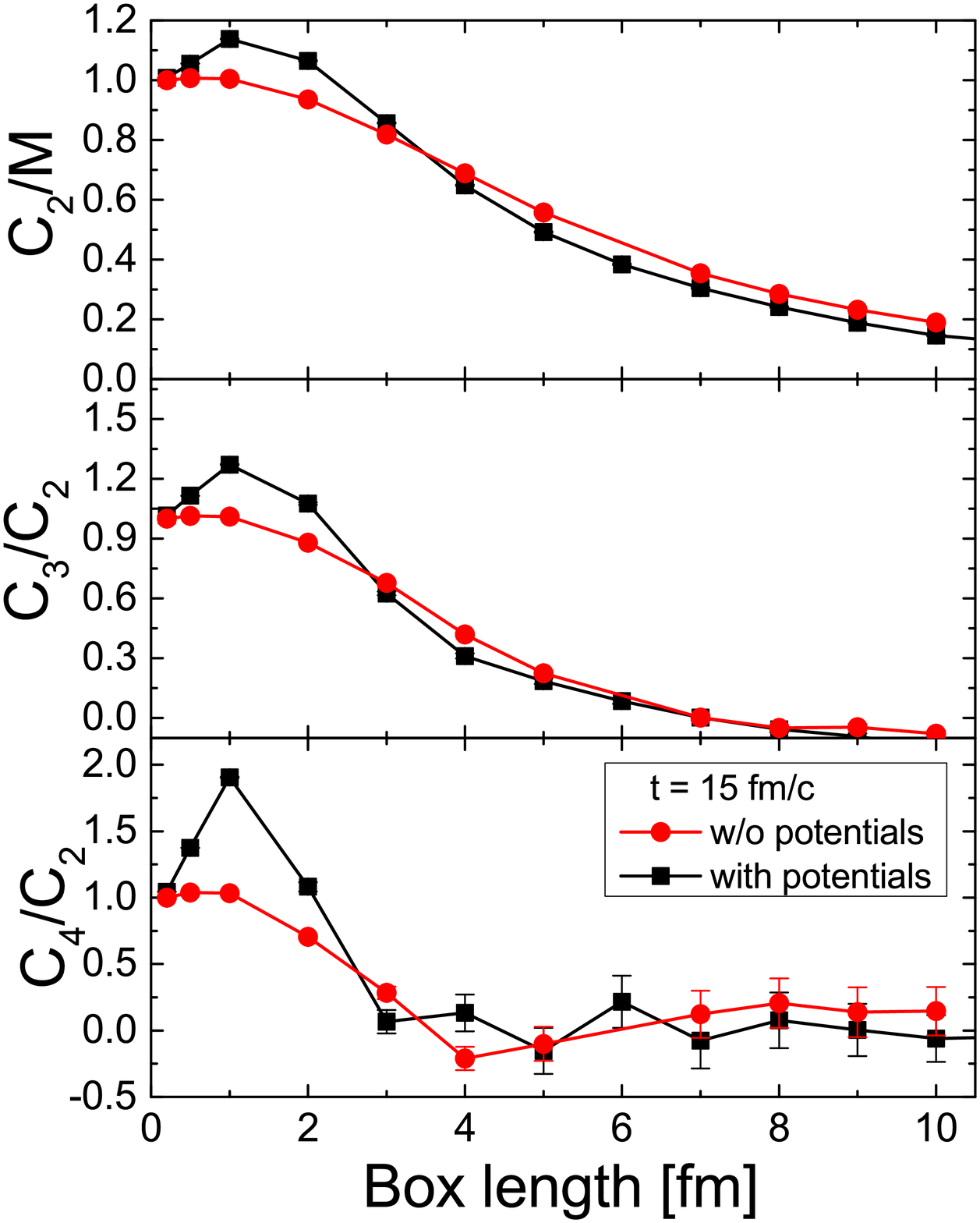}
\caption{[Color online]  Ratios of cumulants of the baryon-number as function of the box length (see figures \ref{f1} and \ref{f2}). Only the results for the early time (t=15 fm/c) are shown, with and without potentials. 
}\label{f3}
\end{figure}		

\section{Results In coordinate space}

\begin{figure}[t]	
\includegraphics[width=0.5\textwidth]{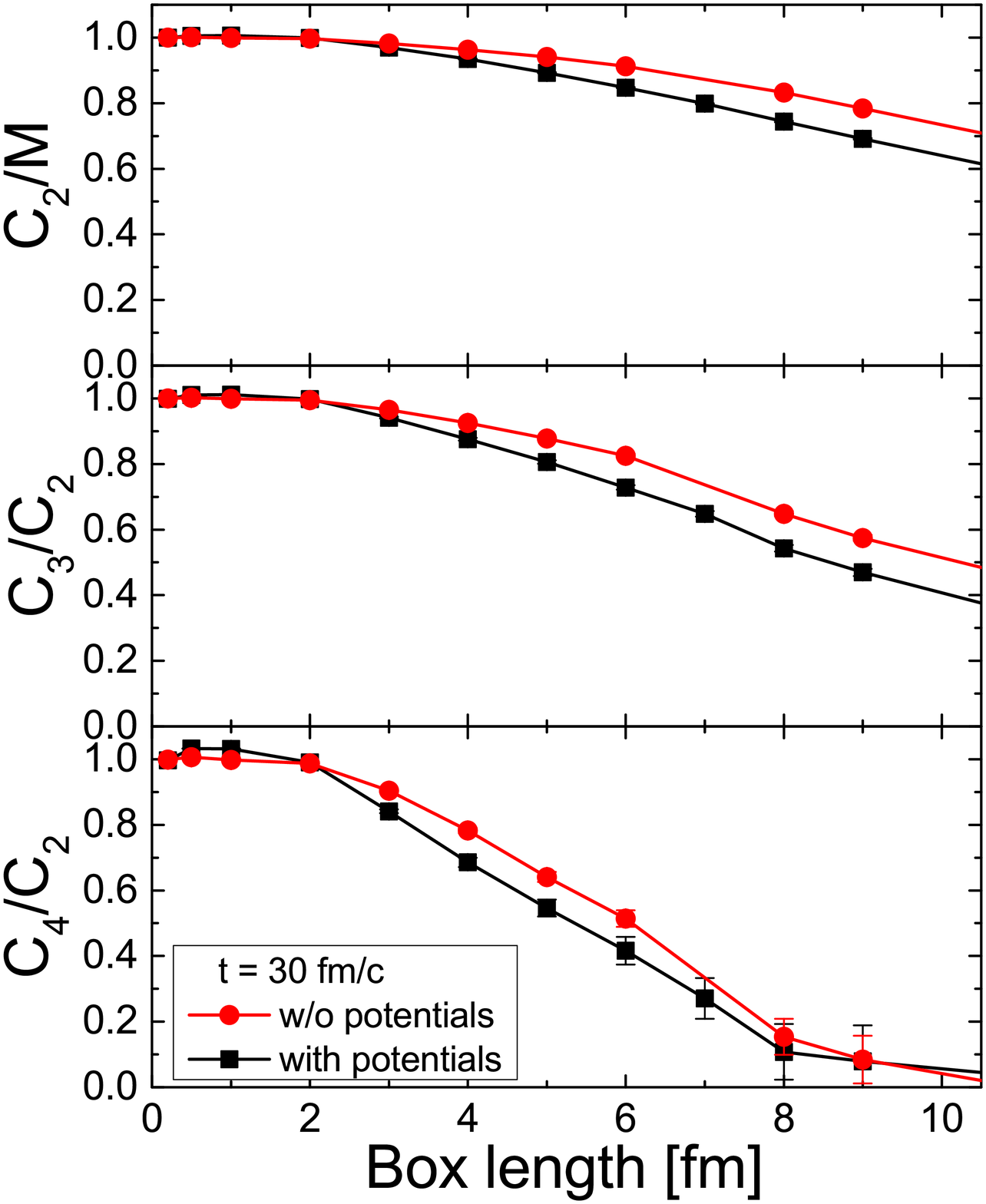}
\caption{[Color online] Ratios of cumulants of the baryon-number as function of the box length (see figures \ref{f1} and \ref{f2}). Only the results for the late time (t=30 fm/c) are shown, with and without potentials.  
}\label{f4}
\end{figure}

Fluctuations and correlations due to a phase transition and critical behavior usually are manifested in coordinate space. For example the spinodal decomposition creates clumps of matter in coordinate space and at the critical endpoint, correlations can extend over large spatial distances. To verify that indeed fluctuations are affected by nuclear interactions, we first have to study the cumulant ratios for a fixed spatial volume, during the dynamical evolution of the system.
Figures \ref{f3} and \ref{f4} show the cumulant ratios, for the net-baryon number, calculated as a function of the box length as defined in the previous section. Again two different times, $t=15$~fm/c and $t=30$~fm/c, are shown. At the early time one can clearly observe an enhancement of all cumulant ratios in the case where nuclear long range interactions are taken into account. The enhancement is strongest for the fourth order cumulant, as expected. The enhancement also only occurs for boxes of length smaller than 2~fm which is due to the finite correlation length. For larger boxes the effect of baryon number conservation begin to dominate and all cumulant ratios decrease.

At the later time the enhancement of the cumulant ratios, in coordinate space, is all but gone. The same systematic enhancement of the cumulant ratios is observed at small box sizes, as for the early times, but the effect is much smaller. This can be understood as a result of the much lower density (sub-saturation density) at the late time. 

\section{Results in momentum space}

\begin{figure}[t]	
\includegraphics[width=0.5\textwidth]{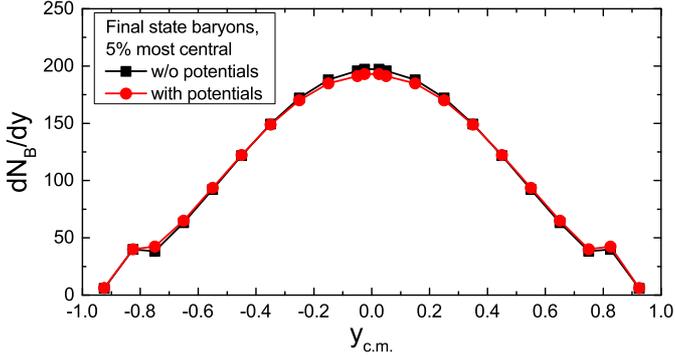}
\caption{[Color online] Final rapidity distribution of baryons in most central collisions ($b<3.4$ fm). Compared are the results for the calculation with and without nuclear potentials. Only a small difference in the mean rapidity distribution is observed.  
}\label{f5}
\end{figure}		

\begin{figure}[t]	
\includegraphics[width=0.5\textwidth]{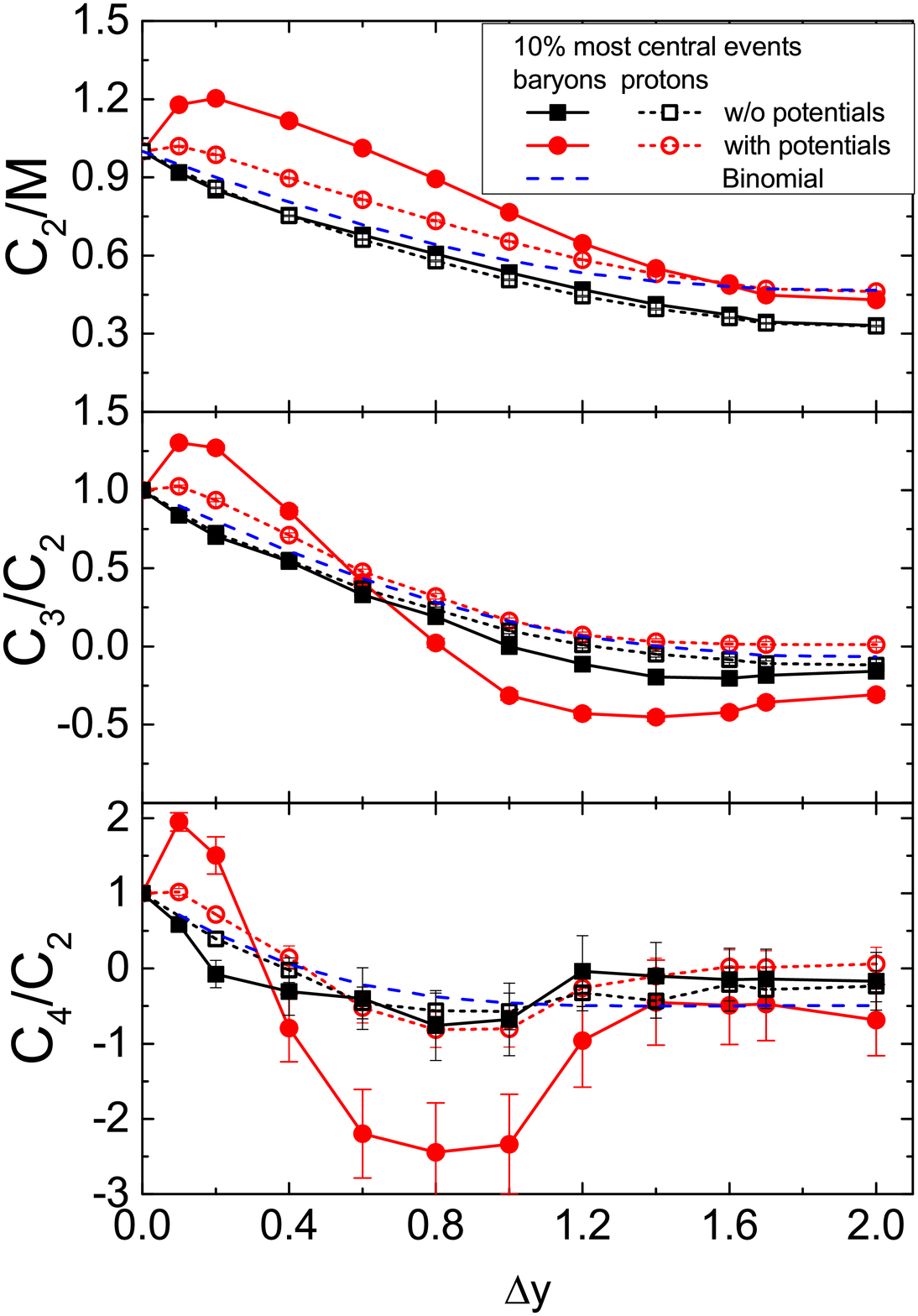}
\caption{[Color online]  Comparison of final proton- and baryon-number cumulant ratios with acceptance cuts, for head on collisions, as function of the rapidity window $\Delta y$, around mid-rapidity. 
}\label{f6}
\end{figure}		

Heavy ion experiments cannot measure coordinate space distributions of baryons during the time evolution of the fireball. They measure momentum space distributions of protons after the final kinetic freeze-out of all particles. 
Furthermore it is not clear that the coordinate space correlations which appear due to the nuclear interactions (or critical phenomena) will translate into momentum space correlation at the end of the systems evolution.

In the following we will therefore present results of our simulations for baryons and protons in the HADES transverse momentum acceptance ($0.4<p_T<1.6$ GeV) and for a given interval in rapidity, around the center of mass rapidity. The results for the cumulant rations will again be for head on ($b=0$~fm) collisions, to avoid strong contributions from volume fluctuations.
Figure \ref{f5} shows the average net-baryon number rapidity distributions for most central Au+Au events. Here we compare cascade mode results with simulations that include nuclear potentials. One can see that the average rapidity distributions are very similar for the two cases, even though the maximum compression varies quite significantly as was shown in figure \ref{f1}.

Finally, figure \ref{f6} shows the results of the net-baryon and net-proton cumulant ratios as function of the rapidity interval $\Delta y$. Here several interesting observations can be made. 

\begin{enumerate}
\item The cumulant ratios for net-baryons are enhanced, in the case where nuclear interactions are enabled, for a small rapidity window $\Delta y < 0.3$.
\item For larger rapidity windows all cumulant ratios are suppressed due to the effect of conservation laws, especially baryon number conservation. 
\item The effect of the enhancement is much smaller for net-protons, as compared to net-baryons, due to the random exchange of iso-spin with neutrons and pions, which are produced abundandly already at this beam energy \cite{Reisdorf:2010ie}. 

\item The cascade mode (black lines with squares) agrees rather well with a simple binomial distribution for the net-baryons. As an input $p$ for the binomial cumulant ratios in equations (\ref{bi}) we simply use the fraction of total baryons in the given rapidity interval. 
\end{enumerate}

\section{Conclusion and Discussion}

We have shown that nuclear interactions can have a significant effect on the net-baryon number cumulant ratios in heavy ion collisions at SIS18 beam energies. This is true for the cumulant ratios in coordinate and momentum space. Furthermore it was shown that an enhancement of the cumulant ratios is only observed for a small acceptance in coordinate or momentum space and that larger acceptance windows are dominated by conservation laws. Finally, we have also shown that the effect is diminished if only net-protons are measured, due to the fact that iso-spin is randomly distributed amongst the baryons. 

Even though the qulitative effect of the nuclear interactions is in agreement with predictions from grand canonical models \cite{Vovchenko:2015pya,Mukherjee:2016nhb}, the quantitative signal is significantly smaller.
This can be explained by the following factors:
\begin{enumerate}
\item The system in heavy ion collisions is small and short lived. Therefore the correlation length is limited not only by the system size but also the short time period the system spends in a dense phase with strong interactions.
\item Coordinate space correlations $\neq$ Momentum space correlations. The increase in the cumulants usually originates from correlations in coordinate space induced by attractive and repulsive interactions. It is not clear that these coordinate correlations completely translate to momentum space correlations which can be measured.
\item Calculations in a grand canonical ensemble do not take into account the conservation of the net baryon number as it occurs in nuclear collisions. In a microscopic transport model this is naturally taking into account.
\end{enumerate}
In conclusion, it was shown that the above discussed factors make it much harder for long range correlations from nuclear interactions or critical behavior to be measured through the proton number cumulants in heavy ion collisions.

\section{Acknowledgments}
The authors thank X.~Luo and Nu~Xu as well as V.~Koch for fruitful discussions.
The computational resources were provided by the LOEWE Frankfurt Center for Scientific Computing (LOEWE-CSC). This work was supported in part by the German Academic Exchange Service (DAAD), the BMBF, the Chinese Scholarship Council, the National Natural Science Foundation of China (Nos. 11505057, 11605270, 11747312, and  11375062), and the Zhejiang Provincial Natural Science Foundation of China (No. LY18A050002) as well as a generous donation by the SAMSON AG. HS acknowledges the support through the Judah M. Eisenberg Laureatus Chair.

\end{document}